\documentclass[twocolumn,runningheads,natbib]{svjour2}
\journalname{Astrophysics and Space Science}
\def\zerozero  {SGR~1900$+$14}
\def\zerosei  {SGR~1806--20}
\def\int {INTEGRAL}
\smartqed  
\usepackage{graphicx}

\begin{document}

\title{Unveiling Soft Gamma-Ray Repeaters with INTEGRAL\thanks{DG acknowledges the French Space Agency (CNES) for financial support. Based on observations with INTEGRAL, an ESA project with instruments and the science data centre funded by ESA member states (especially the PI countries: Denmark, France, Germany, Italy, Switzerland, Spain), Czech Republic and Poland, and with the participation of Russia and the USA. ISGRI has been realized and maintained in flight by CEA-Saclay/DAPNIA with the support of CNES.
KH is grateful for support under NASA's INTEGRAL U.S. Guest Investigator program, Grants NAG5-13738 and NNG05GG35G. }
}


\author{Diego G\"{o}tz         \and
        Sandro Mereghetti \and
	Kevin Hurley 
}

\authorrunning{G\"{o}tz, Mereghetti \& Hurley} 

\institute{D. G\"otz \at
              CEA-Service d'Astrophysique\\
	      Orme des Merisiers, Bat. 709\\
	      91191 Gif-sur-Yvette, France\\
              \email{diego.gotz@cea.fr}           
           \and
           S. Mereghetti \at
              INAF -- Istituto di Astrofisica Spaziale e Fisica Cosmica, Milano, Italy
	      \and
	      K. Hurley \at
	      University of California at Berkeley, Space Sciences Laboratory, Berkeley CA, USA
}

\date{Received: date / Accepted: date}

\maketitle

\begin{abstract}
Thanks to \int's long exposures of the Galactic Plane, the two brightest Soft Gamma-Ray Repeaters, 
SGR 1806-20 and SGR 1900+14, have been monitored and studied in detail for the first time at
hard-X/soft gamma rays.

This has produced a wealth of new scientific results, which we will review here. Since SGR 1806-20 was 
particularly active during the last two years, more than 300 short bursts have been observed with \int~ 
and their characteristics have been studied with unprecedented sensitivity in the 15-200 keV range. 
A hardness-intensity anticorrelation within the bursts has been discovered and the overall 
Number-Intensity distribution of the bursts has been determined. In addition, a particularly active state, during 
which ~100 bursts were emitted in ~10 minutes, has been observed on October 5 2004, indicating that 
the source activity was rapidly increasing. 
This eventually led to the Giant Flare of December 27th 2004, for which a possible 
soft gamma-ray ($>$80 keV) early afterglow has been detected. 

The deep observations allowed us to discover the persistent emission in hard X-rays (20-150 keV)
from 1806-20 and 1900+14, the latter being in a quiescent state, and to directly compare the spectral
characteristics of all Magnetars (two SGRs and three Anomalous X-ray Pulsars) detected with \int.

\keywords{gamma-rays: observations \and pulsars: individual SGR 1806--20, SGR 1900+14 \and pulsars: general}
\PACS{95.85.Pw \and 95.85.Nv \and 96.12.Hg \and 97.60.Gb}
\end{abstract}

\section{Introduction}
\label{intro}
Soft gamma-ray repeaters (SGRs, for a recent review see
\citet{woodsrew}) are a small group (4--7) of peculiar high-energy sources
generally interpreted as ``magnetars'', i.e. strongly magnetised
($B\sim$10$^{15}$ G), slowly rotating ($P\sim$ 5-8 s) neutron
stars powered by the decay of the magnetic field energy, rather than by
rotation \citep{dt92,pac92,td95}. 
They were discovered through
the detection of recurrent short ($\sim$0.1 s) bursts of
high-energy  radiation in the tens to $\sim$hundred keV energy range, with
peak luminosity up to 10$^{39}$-10$^{42}$ erg s$^{-1}$, above the
Eddington limit for neutron stars. The rate of burst emission in
SGRs is highly variable. Bursts are generally emitted during
sporadic periods of activity, lasting days to months, followed by
long ``quiescent'' time intervals (up to years or decades) during
which no bursts are emitted. Occasionally SGRs emit ``giant
flares'', that last up to a few hundred seconds
and have peak luminosity up to  10$^{46}$-10$^{47}$ erg s$^{-1}$.
Only three giant flares have been observed to date, each one from
a different source (see, e.g., \cite{mazets} for 0526--66, \cite{hurley1999} for 1900+14,
\cite{swiftgiant,acsgiant,rhessigiant} for 1806--20).

Persistent (i.e. non-bursting) emission is also observed from SGRs
in the soft X--ray range ($<$10 keV), with a typical luminosity of
$\sim$10$^{35}$ erg s$^{-1}$, and, in three cases, periodic
pulsations with periods of 5 -- 8 seconds have been detected. Such pulsations
proved the neutron star nature of SGRs and allowed the derivation of
spin-down at rates of $\sim$10$^{-10}$ s s$^{-1}$, consistent with
dipole radiation losses for magnetic fields of the order of
B$\sim$10$^{14}$-10$^{15}$ G. The X--ray spectra are generally
described with absorbed power laws, but in some cases strong
evidence has been found for the presence of an additional
blackbody-like component with a typical temperature of $\sim$0.5 keV
\citep{xmm}.

Over the last few years the \int~ satellite \citep{integral}, launched in 2002 and operating in the 15 keV-10 MeV energy range, has provided a wealth of new results concerning
the two brightest SGRs, 1806--20 and 1900+14. Most aspects concerning the SGRs, short bursts, giant flares, and persistent emission, have been investigated, and new results have been found for each of them.
We will review them here. 

\section{\zerosei}
\label{sec:zerosei}
\zerosei~ was  discovered by the Interplanetary Network (IPN) in 1979 \citep{1806disc}. It lies in a crowded region
close to the galactic centre. \citet{kouve1806} discovered a quiescent X-ray pulsating ($P$=7.48 s) counterpart, which
was spinning down rapidly ($\dot P$=2.8$\times$10$^{-11}$ s s$^{-1}$). If this spindown is interpreted
as braking by a magnetic dipole field, its strength is $B\sim$10$^{15}$ G. The source activity is variable, 
alternating between quiet periods and very active ones.

After a period of quiescence, \zerosei~ became active in the Summer of 2003  \citep{hurley2003}.
Its activity then increased in 2004 (see e.g. \citet{moresgr,evenmoresgr}).
A strong outburst during which about one hundred short bursts were emitted in a few minutes
occurred on October 5 2004 \citep{gotz1806}.
Finally a giant flare,
whose energy (a few 10$^{46}$ erg) was two orders of magnitude larger than those of the
previously recorded flares from SGR 0526-66 and SGR 1900+14,
was emitted on December 27$^{th}$ 2004 (see e.g. \citet{swiftgiant,acsgiant,rhessigiant}).

\subsection{Short Bursts}
The results presented in this Section are based on observations obtained with the IBIS coded mask telescope \citep{ibis},
and in particular with its low-energy (15 keV-1 MeV) detector ISGRI \citep{isgri}. More than
400 short bursts have been detected with IBIS/ISGRI. They have been identified either using the triggers provided
by the \int~ Burst Alert System (IBAS, \citet{ibas}), or by computing light curves with 10 ms time 
resolution and looking for significant excesses corresponding to the direction of \zerosei. 

\begin{figure}
\centering
\includegraphics[width=6.2cm]{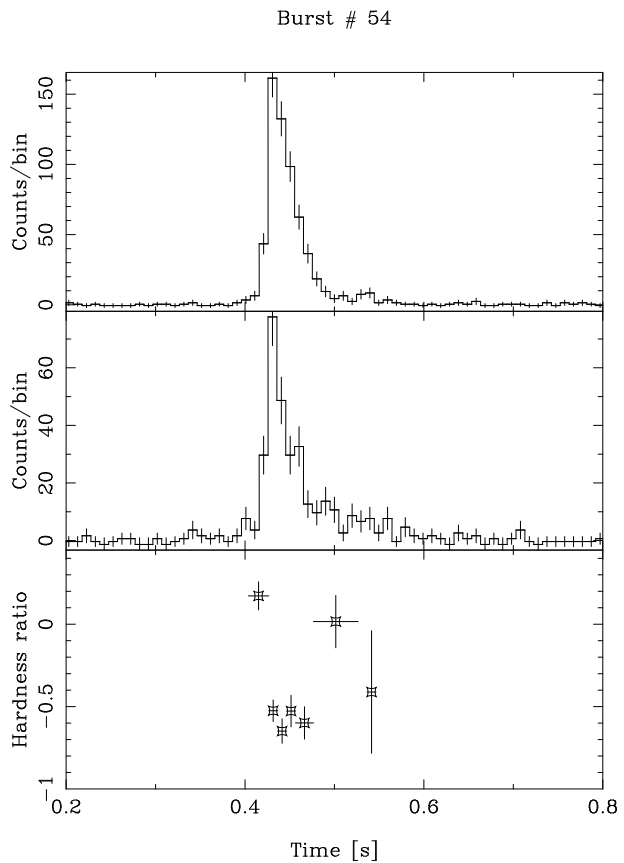}
\caption{IBIS/ISGRI light curves in the soft (20-40 keV, upper
panel) and hard (40-100 keV, middle panel) energy range and
hardness ratio (lower panel) for a short burst from SGR 1806--20.}
\label{fig:evol}      
\end{figure}

\paragraph{Spectral Evolution} All the bursts detected by IBIS
are typical in terms of duration and spectra. The new result provided by the analysis of the \int~ sample is the
spectral evolution within the bursts. By computing time resolved hardness ratios, \citet{gotz04,gotz1806} showed that some bursts evolve significantly with time, especially the ones with a Fast Rise Exponential Decay (FRED) profile.
The hardness ratios have been computed using the background subtracted light curves in two energy bands 
(20-40 ($S$) and 40-100 ($H$) keV) and were defined as $HR$=($H$--$S$)/($H$+$S$).
It turns out that the bursts' peaks tend to be spectrally softer than the bursts' tails. 
This behaviour had been reported earlier 
only for two peculiar bursts originating from \zerozero. These
two bursts were quite different from usual bursts, lasting about 1 s and having a very hard 
spectrum ($kT\sim$ 100 keV, \citet{woods}).
One example of this kind of evolution detected in regular bursts for \zerosei~ is shown in Fig. \ref{fig:evol}.

\begin{figure}
\centering
\includegraphics[width=6cm,angle=-90]{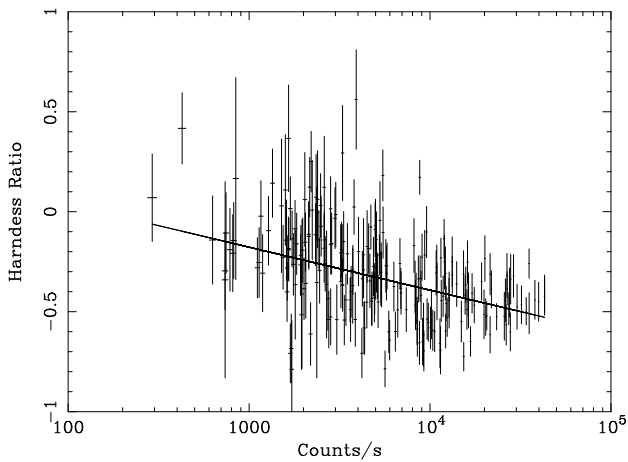}
\caption{Hardness ratio (($H$--$S$)/($H$+$S$)) versus total count rate (20-100 keV,
corrected for vignetting). The points are derived from the time
resolved hardness ratios of the bursts with the best statistics.
The line indicates the best fit with a linear function given in
the text. From \citet{gotz1806}.}
\label{fig:hi}
\end{figure}

The spectral behaviour described above gives rise to a global hardness-intensity anti-correlation.
In fact, by considering all the individual time bins of all the bursts this anti-correlation 
within the bursts has been discovered (see Fig. \ref{fig:hi}). 
To investigate the statistical robustness of the correlation found, the
Spearman rank-order correlation coefficient of the 217 data points, $R_{s}$,
has been computed, which is --0.49. This corresponds to a chance probability of
4$\times$10$^{-15}$ (7.4 $\sigma$) that the distribution is due to uncorrelated data.
According to an F-test the data are significantly (8$\sigma$) better described by a
linear fit ($HR=0.47 - 0.22\times\log I$) than by a constant. This correlation still lacks
of a solid theoretical interpretation since the current Magnetar scenario does not provide a clear
prediction of the burst spectral evolution with time.

\begin{figure}
\centering
\includegraphics[width=8.5cm]{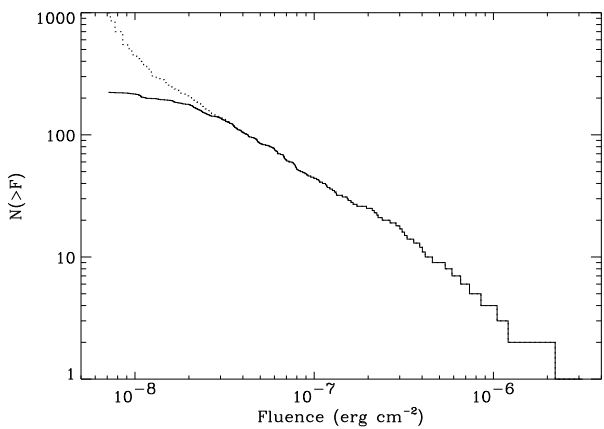}
\caption{Number-intensity distribution of all the bursts detected by \int~ in 2003 and 2004.
The continuous line represents
the experimental data, while the dashed line represents the data corrected
for the exposure. From \citet{gotz1806}.}
\label{fig:logn}
\end{figure}

\paragraph{Number-Intensity Distribution} 
We derived the fluences of 224 bursts using the vignetting and dead-time corrected 
light curves. We applied a conversion factor between counts and physical units derived by the
spectral analysis of the brightest bursts and assuming that the averaged burst spectra do not
change much between bright and faint bursts; for details see \citet{gotz1806}. These fluences
 have been used to compute the number-intensity distribution (Log N-Log S) of the bursts.
The experimental distribution deviates significantly from a single power-law (Fig. \ref{fig:logn}). 
This is first of all due to the fact that the source has been observed at different off-axis angles.
The faintest bursts are missed when the source is observed at large off-axis angles.
In order to correct for this effect we have computed the effective exposure of the source,
taking into account the variation of sensitivity at various off-axis angles.
This yields the exposure-corrected cumulative distribution shown by
the dashed line in Fig. \ref{fig:logn}.

Since the numbers at each flux level are not statistically independent, one cannot use
a simple $\chi^{2}$ minimisation approach to fit the cumulative number-intensity distribution.
So we have used the unbinned detections and applied the Maximum Likelihood
method \citep{crawford}, assuming a single power-law distribution for the number-flux
relation ($N(>S)\propto S^{-\alpha}$). We have
used only the part of the distribution
where completeness was achieved (i.e. $S\geq$3$\times$10$^{-8}$ erg cm$^{-2}$).
In this case the expression to be maximised is
\begin{equation}
{\cal L} = T \ln \alpha - \alpha\sum_{i}\ln S_{i}-T\ln (1-b^{-\alpha})
\end{equation}
where $S_{i}$ are the unbinned fluxes,
$b$ is the ratio between the maximum and minimum values of the
fluxes, and $T$ is the total number of bursts.
This method yields $\alpha$=0.91$\pm$0.09.
If a single power-law model is an adequate representation of data,
the distribution of the quantities
\begin{equation}
y_{i}=\frac{1-S_{i}^{-\alpha}}{1-b^{-\alpha}}
\end{equation}
should be uniform over the range (0,1). In our case, a Kolmogorov-Smirnov (K-S) test
shows that a power law is an appropriate model, yielding a probability of 98.8\% that
the data are well described by our model.

We then divided the bursts into two samples comprising 51 and 173 bursts respectively.
The division is based on the periods of different activity of the source: the 51 bursts were
detected in 1 year and the 173 in 2.5 months.
The two slopes derived with the Maximum Likelihood method
are  $\alpha$=0.9$\pm$0.2 for the low level activity period and
$\alpha$=0.88$\pm$0.11 for the high level one. The two slopes are
statistically consistent with each other and a K-S
test shows that the probability that the two distributions are drawn
from the same parent distribution is 93\%. Thus we conclude that the
the relative fraction of bright and faint bursts is not influenced
by the level of activity of the source.

\begin{figure*}
\centering
\includegraphics[width=10cm,angle=-90]{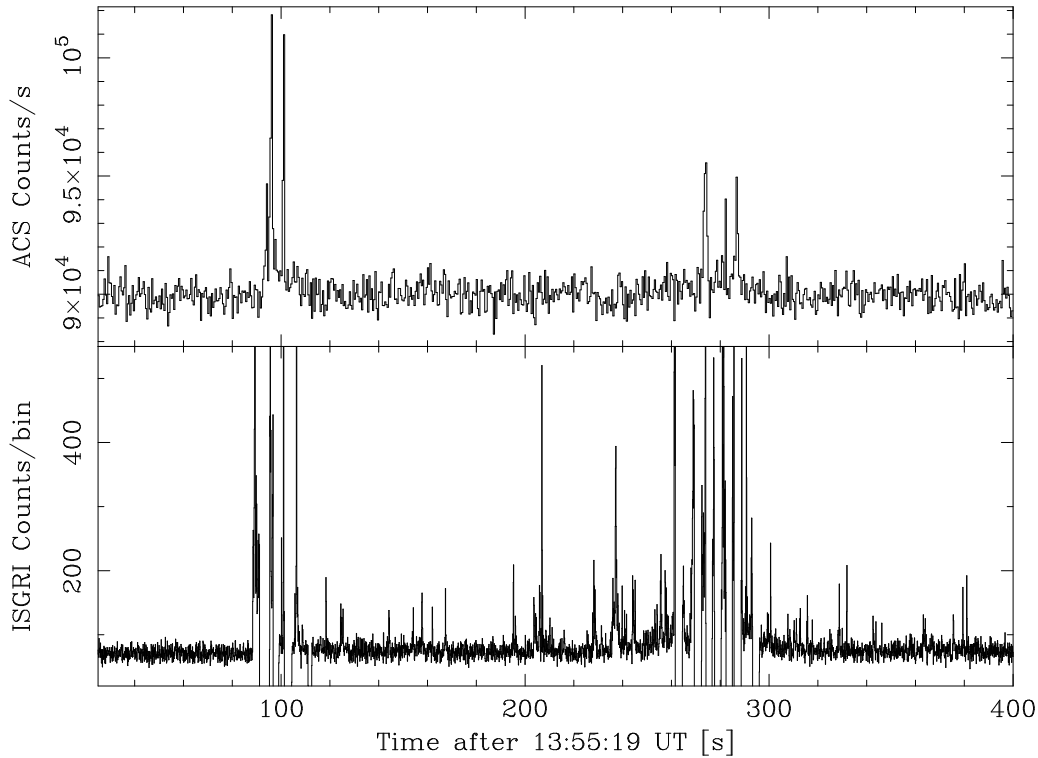}
\caption{Light curves of the initial part of the October 5, 2004 outburst
 of SGR 1806-20. Upper panel: light curve at energy
 greater than $\sim$80 keV obtained with the SPI Anti-Coincidence System in bins of 0.5 s.
Bottom panel: light curve in the 15-200 keV energy range obtained with
  the IBIS/ISGRI instrument (bin size 0.1 s). The gaps in the IBIS/ISGRI light
  curve are due to saturation of the satellite telemetry. From \citet{gotz1806}.}
\label{fig:biglc}
\end{figure*}

\paragraph{The Large Outburst of October 5 2004}
On October 5 2004 IBAS triggered at 13:56:49 UT on a series of bursts originating from \zerosei.
Detailed analysis of this event showed the presence of more than 100 bursts; the activity ended at 14:08:03 UT.
Some bursts were so bright that they saturated the available telemetry share for IBIS, generating some
data gaps lasting up to 10-20 s. The initial part of the outburst is shown in Fig. \ref{fig:biglc}

The fluence of the entire outburst as measured by ISGRI is 1.5$\times$10$^{-5}$ erg cm$^{-2}$, with
a spectrum which is considerably harder than that of the usual short bursts: $kT$=58$\pm$2 keV, using a thermal
bremsstrahlung model.
This fluence value is however heavily affected by the saturation of the brightest bursts and represents only
a lower limit to the real fluence. In order to recover the complete fluence of the event we used the data from the 
Anticoincidence Shield (ACS) of the \int~ spectrometer SPI \citep{spi}. As can be seen in 
Fig. \ref{fig:biglc} (upper panel),
only the brightest bursts are visible in these data and hence they represent
complementary information to the ISGRI data. 

We used the Monte Carlo package MGGPOD \citep{weid05} and a detailed mass modelling of SPI and the whole
satellite (see \citet{weidenspointner} and references therein) to derive the effective area of the ACS
for the direction of \zerosei.
We computed the ACS light curve with a binsize of 0.5 s and estimated the background by 
fitting a constant value to all the data of the same pointing excluding the bursts.
We used the background subtracted light curve to compute the fluence of each burst cluster
in counts. The ACS data do not provide any spectral information, so we computed
the conversion factor to physical units based on the spectral
shapes derived from ISGRI data and on the effective
area computed through our simulations.
The resulting fluences
above 80 keV are 1.2$\times$10$^{-5}$ and 9.4$\times$10$^{-6}$ erg cm$^{-2}$ for the first
and second clusters respectively. Converting these fluences to the 15-100 keV band
one obtains 7.4$\times$10$^{-5}$ and 3.2$\times$10$^{-5}$ erg cm$^{-2}$ respectively. 
By adding these results to the ones obtained for the
ISGRI total spectrum, one can derive the total energy output during the whole event, which is
1.2$\times$10$^{-4}$ erg cm$^{-2}$. This corresponds to 3.25$\times$10$^{42}$ erg
for an assumed distance of 15 kpc \citep{MG05}.

These results can be explained in the framework of a recent
evolution of the magnetar model, where \citet{lyutikov} explains SGR
bursts as generated by loss of magnetic equilibrium in the
magnetosphere, in close analogy to solar flares: new
current-carrying magnetic flux tubes rise continuously into the
magnetosphere, driven by the deformations of the neutron star
crust. This in turn generates an increasingly complicated magnetic
field structure, which at some point becomes unstable to resistive
reconnection. During these reconnection events, some of the
magnetic energy carried by the currents associated with the
magnetic flux tubes is dissipated. The large event described here
can be explained by the simultaneous presence of different active
regions (where the flux emergence is especially active) in the
magnetosphere of the neutron star. In fact, a long outburst with
multiple components is explained as the result of numerous
avalanche-type reconnection events, as reconnection at one point
may trigger reconnection at other points. This explains the fact
that the outburst seems to be composed by the sum of several short
bursts. This kind of event may be correlated with a particularly
complicated magnetic field structure. A large part of the energy stored in the 
manetosphere has then been released during the giant flare on December 27, when
a global restructuring may have taken place. 
This mode also suggests that short events are due to reconnection, while longer
events have in addition a large contribution from the surface,
heated by the precipitating particles, and are harder. This may
explain the generally harder spectra observed. However more ``classical"
scenarios involving only crust fracturing with a large-scale shear
deformation of the crust involving the collective motion of many small units,
without an internal contribution, cannot be ruled out, see e.g. \citet{td01}.

The October 5$^{th}$ event fits the trend of increasing source bursting
activity shown by \zerosei~ in 2003 and 2004. In the same time span also
the luminosity and spectral hardness of the persistent
emission at high (20-150 keV, see below) and low (2-10
keV, \citet{xmm}) energies increased. On the other hand, this peculiar event
did not mark a  peak or a turnover in the SGR activity. In fact
the two XMM observations of  \zerosei~ performed just before
(September 6 2004) and the day after this large outburst (as a ToO
in response to it) yielded similar spectral parameters, fluxes and
pulse profiles, and bursts were seen in both observations \citep{xmm}.

Thus events like these release a small (compared to giant flares)
fraction of the energy stored in the twisted magnetic field of the
neutron star, not allowing the magnetic field to decay
significantly. They are rather related to phases of high activity
due to large crustal deformations (indicating that a large
quantity of energy is still stored in magnetic form) and can be
looked at as precursors of a major reconfiguration of the magnetic
field.

\subsection{The Giant Flare of December 27 2004}
A giant flare from \zerosei~ was
discovered with the \int~ gamma-ray observatory on 2004
December 27 \citep{flare}, and detected with many other satellites (e.g. \citet{swiftgiant,rhessigiant})
The analysis of the SPI-ACS data ($>$80 keV) of the flare, presented in \citet{acsgiant}, show that
the giant flare is composed by an initial spike lasting 0.2 s followed by a $\sim$400 s long pulsating tail,
modulated at the neutron star period of 7.56 s. 
The initial spike was so bright that it saturated the ACS, so we could derive only a lower limit on its fluence,
which turned out to be two orders of magnitude brighter (10$^{46}$ ergs, see e.g. \citet{tera}) 
than the previously observed giant flares from
\zerozero~ \citep{hurley1999}, and SGR 0526--66 \citep{mazets}. The energy contained in the tail (1.6$\times$10$^{44}$ 
ergs), on the other hand, was of the same order as the one in the pulsating tails of the previously observed giant 
flares. 

A $\sim$0.2 s long small burst was detected in the ACS data 2.8 s after the initial spike. It is superposed on the 
pulsating tail and has no clear association with the pulse phase. This burst has been interpreted by \citet{acsgiant} 
as the reflection by the Moon of the initial spike of the giant flare. In fact this delay corresponds to the light 
travel time between \int, the Moon, and back. A similar detection was reported with the Helicon-Coronas-F 
satellite \citep{mazets05}.

\begin{figure*}
\centering
\includegraphics[width=13cm]{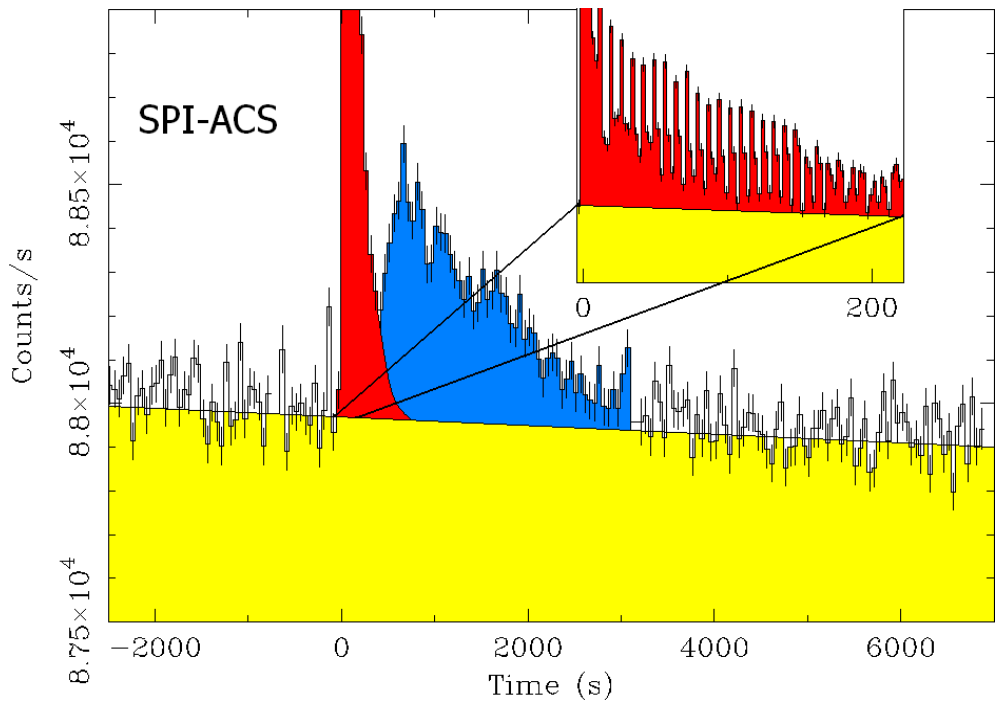}
\caption{Light curve of the Giant Flare of December 27 2004 as measured with 
SPI-ACS above 80 keV. The light curve is binned at 50 s, and hence the pulsating tail
is not visible (it is visible in the inset where the light curve is binned at 2.5 s).
(yellow: instrumental background, red: Flare tail, blue: high-energy afterglow, see text)}
\label{fig:flare}
\end{figure*}

The most striking feature provided by the \int~ data is the detection of a possbile early high-energy afterglow 
emission associated with the giant flare. At the end of the pulsating tail the count rate increased again, forming 
a long bump which peaked around t$\sim$700 s and returned to the pre-flare background level at t$\sim$3000-4000. This
component decays as $\sim t^{-0.85}$, and is shown in blue in Fig. \ref{fig:flare}, while the overall long term background trend is shown in yellow,
and the giant flare itself in red. The association of this emission with \zerosei~ is discussed
in \citet{acsgiant}. The fluence contained in the 400-4000 s time interval is approximately the same as that in the
pulsating tail. With simple gamma-ray burst afterglow models based on synchrotron emission one can derive
the bulk Lorentz factor $\Gamma$ from the time $t_{0}$ of the afterglow 
onset: $\Gamma\sim$15($E$/5$\times$ 10$^{43}$ ergs)$^{1/8}$($n$/0.1 cm$^{-3}$)$^{-1/8}$($t_{0}$/100)$^{-3/8}$,  
where $n$ is the ambient density. This is consistent with the mildly relativistic outflow inferred from the radio 
data \citep{granot}.

\subsection{Discovery of the persistent emission}
In 2005 two groups reported independently the discovery of persistent hard X-ray emission originating from \zerosei~ 
\citep{mere05,molkov}. Up to then, spectral information on the perisistent emission
of SGRs was known only below 10 keV. The low energy spectrum is usually well described by 
the sum of a power law component and a black body.

\begin{figure}
\centering
\includegraphics[width=5.3cm,angle=-90]{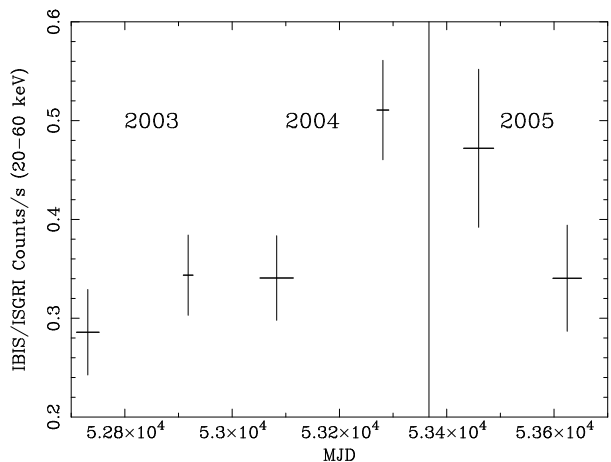}
\caption{Long term light curve of \zerosei, as measured with IBIS. The vertical 
line represents the time of the giant flare of December 27 2004.}
\label{fig:zerosei}
\end{figure}

The spectrum above 20 keV is rather hard with a photon index between 1.5 and 2.0 and extends up to 150 keV 
without an apparent cutoff. It connects rather well with the low energy ($<$ 10 keV) spectrum \citep{xmm}, 
and the intensity and spectral hardness are correlated with the degree of bursting activity of 
the source \citep{mere05,gotz1806} and with the IR flux \citep{giallo}. 
Our group is continuosly monitoring the hard X-ray flux of \zerosei, and the 
long term light curve of the source is shown in Fig. \ref{fig:zerosei}. As can be seen, the persistent 
flux increased in 2003 and 2004 up to the giant flare (which is marked with a vertical line in the plot), and 
then decreased in 2005.

This behaviour can be interpreted as an increase of the twist angle in the magnetar magnetic field, 
which in turn increases the burst emission rate, and produces harder spectra, as predicted by \citet{tlk}.

\section{\zerozero}
\zerozero~ was discovered in 1979 by \citet{1900discovery} when it emitted 3 bursts in 2 days.
Since then short bursts were observed from this source with BATSE, RXTE and
Interplanetary Network satellites in the years 1979-2002. \zerozero~ emitted a
giant flare on August 27 1998 (e.g. \citet{hurley1999}), followed by less intense
``intermediate'' flares on August 29 1998 \citep{ibrahim} and in April 2001 \citep{lenters}. The last bursts
reported from \zerozero~ were observed with the Third Interplanetay
Network (IPN) in November 2002 \citep{ipn1900}. No bursts from
this source were revealed in all the \int~ observations from
2003 to 2005, but Swift has detected renewed activity in 2006 \citep{swift1900}.

\subsection{Discovery of the persistent emission}

Using 2.5 Ms of \int~ data, \citet{gotz1900} reported the discovery of persistent hard X-ray emission, this time
from a quiescent SGR, 1900+14. This emission extended up to $\sim$ 100 keV, but with a softer spectrum 
compared to \zerosei, having a photon index of 3.1$\pm$0.5. Also the luminosity is dimmer in this case, being
$\sim$4$\times$10$^{35}$ erg s$^{-1}$, a factor of three lower than \zerosei. The \int~ observations spanned 
March 2003 to June 2004, and did not include the recent reactivation of the source in March 2006 
\citep{swift1900}, when the source emitted a few tens of regular bursts plus an intense burst series, 
lasting $\sim$30 s \citep{1900outburst}, reminiscent of the October 5 2004 event from \zerosei.
We recently analysed the \int~ data spanning from August 2004 to March 2006, and found that the
hard X-ray flux of the source flux did not increase up to a few weeks before its reactivation.
This indicates that the reactivation was not triggered by a flux increase, at least on the time
scale of a few months sampled by \int.

The soft and constant spectrum of \zerozero ~is possibly related to the fact that
this source is still in a rather quiescent state.

\section{Comparison with the Anomalous X-ray Pulsars}

Hard X-ray persistent emission ($>$20 keV) has recently been
detected from another group of sources,  the Anomalous X-ray
Pulsars (AXPs, \citet{axps}), which share several
characteristics with the SGRs and are also believed to be
magnetars (see \citet{woodsrew}).
Hard X-ray emission has been detected from three AXPs with
\int: 1E 1841--045 \citep{molkovaxp}, 4U 0142+61
\citep{denhartog} and 1RXS J170849--400910 \citep{revnitsev}.
The presence of pulsations seen with RXTE up to $\sim$200 keV in 1E
1841--045 \citep{kuiper} proves that the hard X-ray emission
originates from the AXP and not from the associated supernova
remnant Kes 73. The discovery of (pulsed) persistent hard X-ray
tails in these three sources was quite unexpected, since below 10
keV the AXP have soft spectra, consisting of a blackbody-like
component (kT$\sim$0.5 keV) and a steep power law (photon index
$\sim$3--4).

\begin{figure}
\centering
\includegraphics[width=8.8cm]{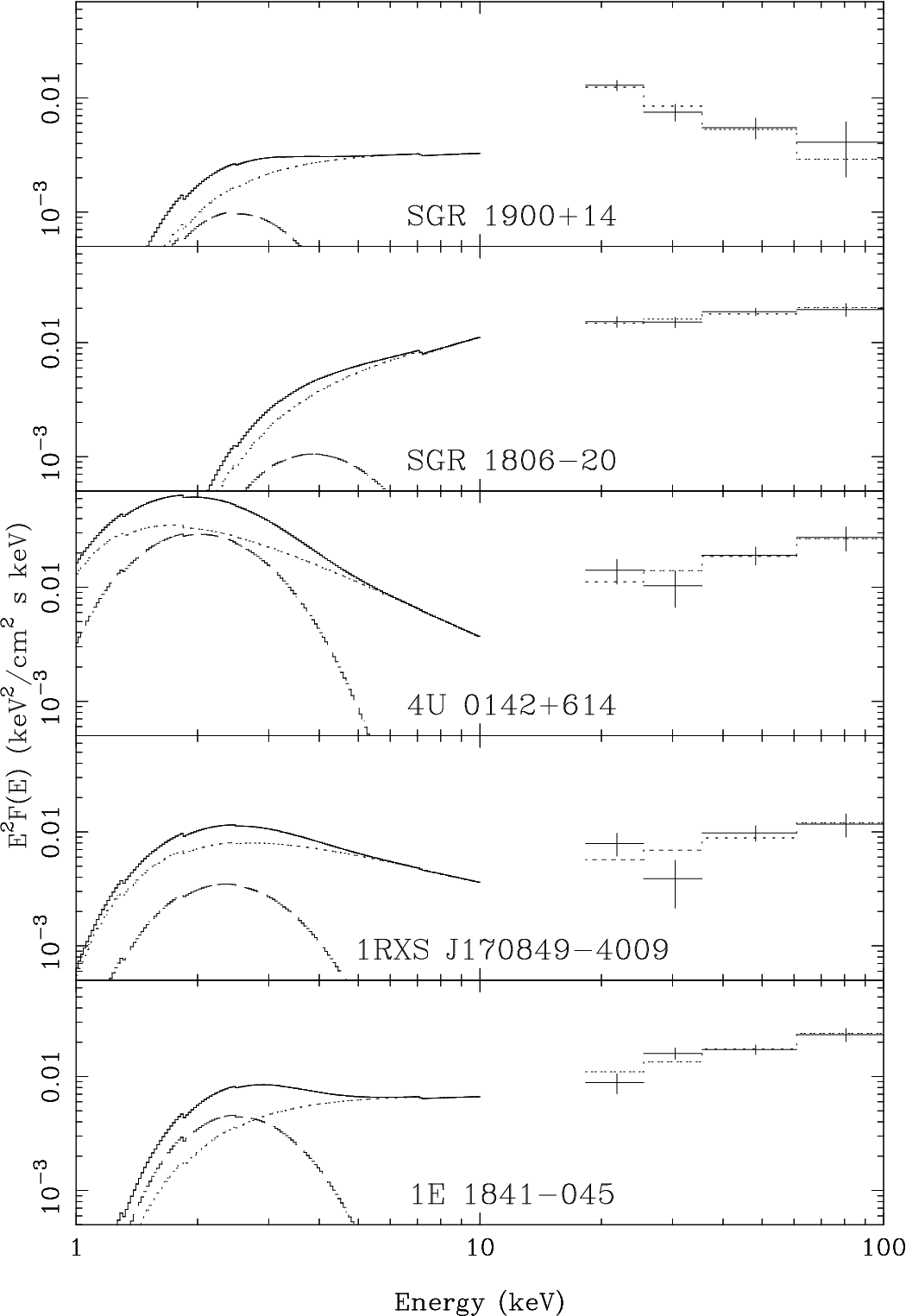} 
\caption{Broad band X--ray spectra of the five magnetars detected by \int~. The
data points above 18 keV are the \int~ spectra with their best fit
power-law models (dotted lines). The solid lines below 10 keV represent
the absorbed power-law (dotted lines) plus blackbody (dashed lines) models
taken from \citet{woods01} (\zerozero , during a quiescent state in spring 2000), \citet{xmm} (\zerosei, observation B, when the bursting activity was low), \citet{gohler} (4U~0142+614), \citet{rea}
(1RXS~J170849--4009), and \citet{morii} (1E~1841--045). From \citet{gotz1900}.}
\label{fig:bbsp}
\end{figure}

In order to coherently compare the broad band spectral properties
of all the SGRs and AXPs detected at high energy, we analysed all
the public \int~ data using the same procedures. Our results are shown
in Fig. \ref{fig:bbsp}, where the \int~ spectra are plotted together
with the results of observations at lower energy taken from the
literature (see figure caption for details).

As can be seen, AXPs generally present harder spectra than SGRs in hard X-rays. In particular, for the three
AXPs, a spectral break is expected to occur between 10 and 20 keV in order to reconcile the soft and the 
hard parts of the spectrum. On the other hand SGRs, present a softer spectrum at higher energies also implying
a break around 15 keV (especially for \zerozero) but in the opposite sense with respect to the AXPs. The fact that
the spectral break is more evident in \zerozero~ could be due to the fact that its level of activity was much lower
during our observations, compared to \zerosei. All three AXPs, on the other hand, can be considered to have been in a quiescent
state since no bursts were been reported from them during the \int~ observation.

The magnetar model, in its different flavours, explains this hard X-ray emission as powered by brems\-strah\-lung photons
produced either close to the neutron star surface, or at a high altitude ($\sim$ 100 km) in the magnetosphere
\citep{tlk,tb05}.  The two models can be distinguished by the presence of a cutoff at $\sim$ 100 keV or $\sim$ 1 MeV. 
Unfortunately current experiments like \int~  are not sensitive enough to firmly assess the
presence of the cutoffs and hence to distinguish between the two models.

\section{Conclusions}

Thanks to \int, and in particular to its imager IBIS, we have been able to study most of the magnetars' phenomenology
with unprecedented sensitivity at high energies. One of the most striking results is the discovery, which
was particularly unexpected for AXPs, of the persistent hard X-ray emission. This discovery, which
can be considered one of the most important \int~ results at all, represents a new important input for theoreticians
who started to include it in the magnetar model (see e.g. \citet{belo}). 

Also, the fact that short bursts evolve with time is a new feature that has to be considered with care within the 
magnetar model: up to now no clear explanation has been provided for this.

The large number of detected short bursts from \zerosei~ allowed us a good
determination of the shape and slope of their Number-Intensity distribution, showing that a single
power law holds over 2.5 orders of magnitude. 

In addition, the fact that \zerosei~ has been particularly active in these last years, also emitting a 
once-in-a-lifetime event such as the giant flare (and its possible high energy afterglow), has allowed  
observations of relatively rapid changes of the bursting and persistent emission of a Magnetar and   
to interpret then with the evolution of a very strong and complicated magnetic field, confirming
the magnetic field as the dominant source of energy in Soft Gamma-Ray Repeaters and Anomalous X-ray
Pulsars.

\bibliographystyle{spmpsci}


\end{document}